\documentclass[12pt,journal,final,oneside,twocolumn,nofonttune]{IEEEtran}
\usepackage[compress]{cite}
\usepackage{hyperref}
\makeatletter
\renewcommand{\citepunct}{,\penalty\@m\hskip.13emplus.1emminus.1em}
\renewcommand{\citedash}{\hbox{--}\penalty\@m}
\makeatother

\usepackage{graphicx}
\usepackage{amsmath}
\usepackage{amssymb}
\usepackage{amsfonts}
\usepackage{psfrag}
\usepackage{float}
\usepackage{array}
\ifCLASSOPTIONcompsoc
\usepackage[tight,normalsize,sf,SF]{subfigure}
\else
\usepackage[tight,footnotesize]{subfigure}
\fi
\IEEEoverridecommandlockouts
\begin{document}
\title{Joint Channel Direction Information Quantization For Spatially Correlated 3D MIMO Channels}
\author{
\IEEEauthorblockN{\large{Fang Yuan, Chenyang Yang, Yang Song, Lan
Chen, Yuichi Kakishima and Huiling Jiang}}

\thanks{Fang Yuan  and Chenyang Yang are with
the School of Electronics and Information Engineering, Beihang
University, Beijing China (email: yuanfang@ee.buaa.edu.cn,
cyyang@buaa.edu.cn).

Yang Song  and Huiling Jiang are with DOCOMO Beijing Communications
Laboratories Co., Ltd
(email:\{song,jiang\}@docomolabs-beijing.com.cn). Lan Chen and
Yuichi Kakishima are with NTT DOCOMO, INC. (email:
chenl@nttdocomo.co.jp, yuichi.kakishima.vc@nttdocomo.com). } }

\maketitle

\begin{abstract}
This paper proposes a codebook for jointly quantizing channel
direction information (CDI) of spatially correlated
three-dimensional (3D) multi-input-multi-output (MIMO) channels. To
reduce the dimension for quantizing the CDI of large antenna arrays,
we introduce a special structure to the codewords by using Tucker
decomposition to exploit the unique features of 3D MIMO channels.
Specifically, the codeword consists of four parts each with low
dimension individually targeting at a different type of information:
statistical CDIs in horizontal direction and in vertical direction,
statistical power coupling, and instantaneous CDI. The proposed
codebook avoids the redundancy led by existing independent CDI
quantization. Analytical results provide a sufficient condition on
3D MIMO channels to show that the proposed codebook can achieve the
same quantization performance as the well-known rotated codebook
applied to the global channel CDI, but with significant reduction in
the required statistical channel information. Simulation results
validate our analysis and demonstrate that the proposed joint CDI
quantization provides substantial performance gain over independent
CDI quantization.
\end{abstract}

\section{Introduction}
To meet the ever-growing data demand in future 5th Generation (5G)
cellular networks, one of the promising ways is to increase the
number of antennas at the base station (BS) \cite{Com2014}. However,
equipping large number of antennas at a BS is challenging due to the
physical space limitation. This naturally calls for
three-dimensional (3D) multi-input-and-multi-output (MIMO) systems
\cite{Nam2013,DOCOMO,ALSB2011,Rappaport2013}, where
active antenna elements are placed in a two-dimensional (2D) array, e.g.,
rectangular and circular arrays, or even non-planar arrays, at the BS.

The non-linear arrays can provide both horizonal and vertical spatial
resolution, and thus can support 3D beamforming \cite{DOCOMO}. 3D
beamforming can be user-specific, which improves the signal to
noise ratio (SNR) meanwhile generating less
interference to adjacent users. Recently, a 3D MIMO prototype system
operating at millimeter-wave bands was reported to support a
multi-Gbps data rate service in macro cells, which provides high array
gain to compensate the severe path loss \cite{Rappaport2013}.

In practice, the promised performance of 3D beamforming largely
depends on how accurate the channel direction information (CDI) is
obtained at the BS. In time division duplexing (TDD) systems, the CDI obtained by channel estimation in uplink can
be used for beamforming in downlink if the antennas
at the BS are perfectly calibrated
\cite{Com2014}, where the performance is limited by pilot
contamination \cite{Marzetta2011}. In frequency division duplexing
(FDD) systems, limited feedback is widely used, where the CDI is
firstly quantized at the user and then fed back to the BS
\cite{Love2003}. Yet the feedback overhead is supposed to increase
with the number of antennas \cite{Jindal2006}, which is not acceptable for large
antenna array systems.

Spatial correlation is observed very typical in MIMO channels
\cite{WINNER2}, due to small spacing between adjacent antennas and
low angular spreads. In \cite{Clerckx2008}, B. Clerckx, \emph{et
al.} found that spatial correlation can be exploited to reduce the
overhead for feeding back CDI significantly. The same conclusion was
drawn in \cite{Adhikary2013,Kuo2012,Choi2014} for spatially
correlated massive MIMO channels, which indicates that FDD is also
applicable for large antenna array systems without heavy feedback
overhead as supposed to be.

Various codebooks have been proposed for spatially correlated
channels. Theoretically, Llyod algorithm \cite{Gray} can be applied
to generate codebooks for 3D MIMO channels, but they are difficult
to be used off-line in practice. There are some codebooks obtained
from Llyod algorithms  with reduced complexity, e.g., local-packing
codebook \cite{Raghavan07}, and gain and phase separated
quantization \cite{Huang2011}.  A well-known codebook, rotated
codebook \cite{Love2006}, transforms the codewords optimized for
uncorrelated channels (e.g., Grassmannian subspace packing (GSP)
codewords) by channel correlation matrix. It was proved that the
rotated codebook is asymptotically optimal in quantizing any
spatially correlated channels as the codebook size becomes large
\cite{Rao2006}. An extension of rotated codebook to multiuser MIMO
system is  provided in \cite{Choi2013a}, where not only each user's
own but also the other users' correlation matrices are employed for
the codeword rotation. When designing codebooks for real-world
cellular systems, practical limitations such as constant modulus and
finite alphabet need to be taken into account. Discrete Fourier
transformation (DFT) codebook meets these limitations, which is
suitable for highly correlated channels with uniform linear array
 at the BS \cite{Yang2010}. For 3D MIMO system with
uniform rectangular array (URA), a Kronecker-product DFT
codebook was proposed in \cite{Xie2013}. However, as shown in
\cite{Yang2010}, the performance of these pure DFT based codebooks
degrades severely when the angular spreads of channel increase.

Generally speaking, there are two straightforward strategies to
extend existing codebooks to 3D MIMO systems. The first strategy is
\emph{global channel quantization}, which expresses the 3D MIMO
channel as a larger global vector, and reuse existing codebooks
proposed for 2D MIMO channels. For example, the codebooks in
\cite{Love2006,Raghavan07,Huang2011,Choi20122} can be applied to
quantize the global 3D MIMO CDI directly. However, due to the high
dimension of the resulting CDI vector, it is challenging for the BS
to obtain accurate channel statistical information required by the
global channel quantization to reuse existing codebooks, e.g.,
channel correlation matrix used by the rotated codebook, which is
known as the ``curse of the dimensionality'' \cite{Marimont1978}.
Moreover, large dimension incurs high complexity in matrix
operations \cite{Marzetta2011a}. The second strategy is
\emph{independent CDI quantization}, which quantizes the CDI of 3D
MIMO in horizontal and vertical directions independently and reuses
existing codebooks in each direction  \cite{ALSB2011}. This strategy
is simple and avoids the problem caused by the high dimensional
channel vector, which however comes at a cost of low quantization
accuracy when the angular spread is large, as will be clarified
later.

To design a desirable codebook in 3D MIMO channels under various
angular spreads and reduce the amount of required spatial
correlation information, we propose a joint CDI quantization
strategy to quantize the CDI of two directions jointly. The designed
codeword has a special structure led by the Tucker decomposition
\cite{Stefan2012}, which exploits the unique features of 3D MIMO
channels.

The first feature of 3D MIMO channel is the inherent geometrical
structure inherited from the regularity of antenna arrays. For
example, the array response of 3D MIMO with URA for each ray in the
channel can be decomposed into two subarray responses of lower
channel dimensions in horizontal and vertical directions. Similar
decompositions can be also found for those large arrays nested from
several smaller identical subarrays. This feature can be exploited
for coping with the high-dimension quantization problem in large
antenna arrays.

The second feature is power coupling. When the array responses of 3D
MIMO channels are decomposed into two subarray responses
respectively in horizontal and vertical directions, a common ray
gain is shared by the two subarray responses and not decomposable.
The power coupling is important in the codebook design, since it
connects with the subarray responses from different lower channel
dimensions. The feature can be exploited to avoid the redundancy in
the CDI quantization.

Although many research efforts have been made for the
dimension reduction problem, they mainly employ the
singular value decomposition (SVD) of correlation matrix to exploit the
channel correlation, e.g., \cite{Ko2009}. As far as the
authors known, the unique
features in 3D MIMO channels are not yet
exploited. In this paper, we propose a codebook
for jointly quantizing the CDIs in horizontal and vertical
directions, where a new codeword structure is introduced to reduce the dimension by using the features of 3D MIMO channels.
The contributions are two-fold:
\begin{enumerate}
\item A joint CDI quantization codebook is proposed to quantize spatially correlated 3D MIMO
channels, which has a special codeword structure. The spatial
correlation information required by the cobook for general antenna arrays
is provided  by using the Tucker decomposition. We show that the proposed codebook provides
better quantization performance than the independent CDI
quantization, and requires much less spatial correlation information
than the globally rotated codebook.
\item The performance of joint CDI quantization codebook is analyzed.
We show that the performance is the same as the globally rotated
codebook for 3D MIMO systems with URA under the channels with weak identically independent
distributed (i.i.d.) rays.
\end{enumerate}

The rest of this paper is organized as follows.  In Section II, we present the
3D MIMO channel model and channel representations. In Section III, we propose the
joint CDI quantization codebook for spatially correlated channels.
In Section IV, a solution to find the spatial
correlation information for general antenna arrays
is provided  by using the Tucker decomposition. The performance of the joint CDI quantization
codebook is analyzed in Section V. Simulation results are provided
in Section VI and the paper is concluded in Section VII.

\emph{Notations}: $(\cdot)^T$, $(\cdot)^H$ and $(\cdot)^\ast$ are
respectively the transpose, Hermitian and conjugate operation,
$\odot$ and $\otimes$ are respectively the element-wise and
Kronecker matrix product, $|\cdot|$ and $\|\cdot\|$ are respectively
the absolute value and norm, $\pmb{E}\{\}$ means the expectation
operation, $\text{diag}(\pmb{x})$ is the diagonal matrix with
diagonal entries given in the vector $\pmb{x}$, and
$\text{diag}(\pmb{X})$ is the diagonal matrix with the diagonal
equal to that of the matrix $\pmb{X}$.

\section{System and Channel Models}
\begin{figure}[t]
   \centering
   \includegraphics[width=3.5in]{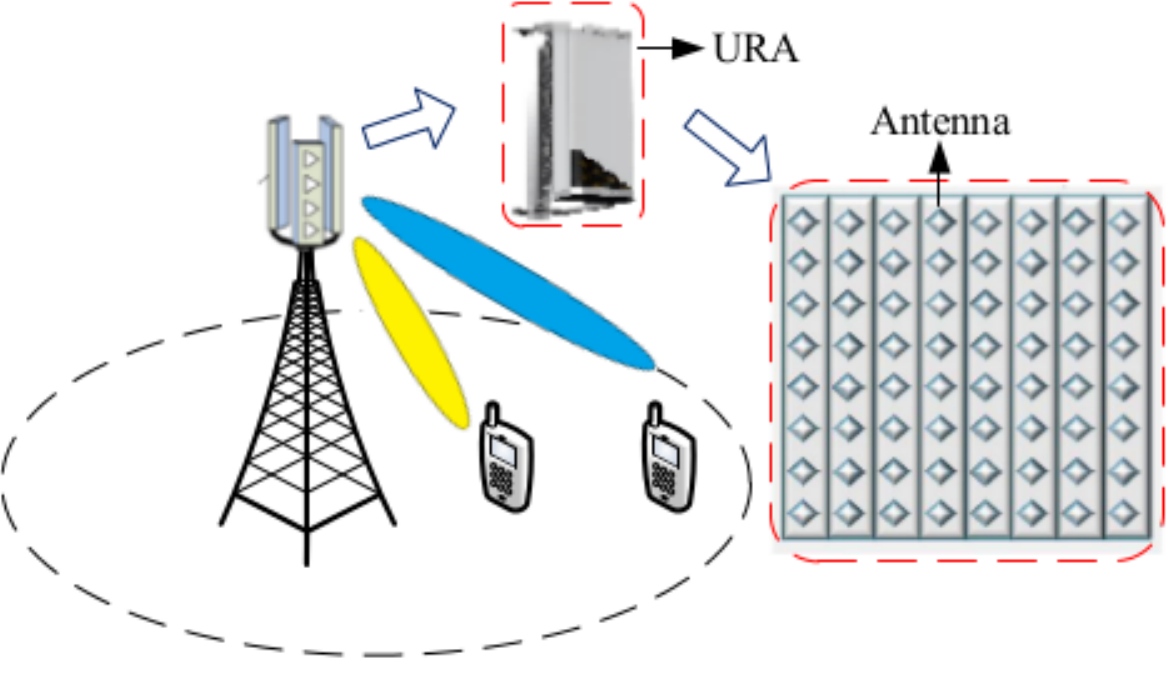}
   \caption{An example of 3D  MIMO system with $8\times 8$ URA at the BS.}\label{3dmf}
\end{figure}

Consider a downlink 3D MIMO system where  an array with $N_t$  antennas is mounted at the BS \cite{Rajagopal2011,Nam2013}, and all antennas are omni-directional. An example of 3D
MIMO system with URA is shown in Fig. \ref{3dmf}.

For easy exposition, we start by considering the 3D multiple-input-and-single-output (MISO)
system with single-antenna users,
and then extend the design to multi-antenna
users. For simplicity, we will not distinguish 3D MISO
channel from 3D MIMO channel hereafter.

According to the generic spatial channel modeling methods
\cite{Yong2003,WINNER2,rel12cm}, the 3D MIMO channel consists of
several scattering clusters distributed in the 3D space, where in
each cluster there are multiple rays with small random angle
offsets, which can be expressed as a $N_t$-dimensional vector
\begin{align}
\pmb{h} &\triangleq \sum\nolimits_{n=1}^N\sum\nolimits_{m=1}^M
g_{n,m}\pmb{a}(\Theta_{n,m})\label{CNM}
\end{align}
where $g_{n,m}\in \mathbb{C}$ is the random gain of the
$m$th ray in the $n$th cluster with zero mean, the angle
vector $\Theta_{n,m}=[\theta_{n,m},\phi_{n,m}]^T$ is specified by
the angle coordinates $\theta_{n,m}$ and $\phi_{n,m}$ in the 3D
space, and $\pmb{a}(\Theta_{n,m})\in \mathbb{C}^{N_t\times1}$ is the
corresponding array response.

The expression of array response $\pmb{a}(\Theta_{n,m})$
depends on the specific form of the array mounted at the BS. The
array responses of some well-structured arrays can be decomposed
into the Kronecker product of subarray responses, while the others
may not. Usually, the large antenna arrays nested from smaller
identical subarrays are decomposable. For example, as shown in
\cite{Yong2003,VT02}, the array response of URA can be decomposed
into two subarray responses respectively in horizontal and vertical
directions as
\begin{align}
\pmb{a}(\Theta_{n,m})=\pmb{a}_{v}(\phi_{n,m}) \otimes
\pmb{a}_{h}(\theta_{n,m})\label{decopm}
\end{align}
with
\begin{align}
\pmb{a}_{h}(\theta_{n,m})\!&=\! [1, e^{j2\pi
\frac{d_h}{\lambda}{\cos\theta_{n,m}}},\!\cdots\!, \!e^{j2\pi
\frac{d_h}{\lambda}(N_h-1){\cos\theta_{n,m}}}]^T\nonumber\\
\pmb{a}_{v}(\phi_{n,m})\!&= \![1, e^{j2\pi
\frac{d_v}{\lambda}{\cos\phi_{n,m}}},\!\cdots\!,\! e^{j2\pi
\frac{d_v}{\lambda}(N_v-1){\cos\phi_{n,m}}}]^T\nonumber
\end{align}
where $d_h$ and $d_v$ are respectively the antenna spacing in
horizontal and vertical directions, $N_h$ and $N_v$ are the number
of antennas at the URA in horizontal and vertical directions,
$\lambda$ is the carrier wavelength, $\cos\theta_{n,m}$ and
$\cos\phi_{n,m}$ are the direction cosines of the $m$th ray in the
$n$th cluster respectively in the horizontal and vertical
directions.

One example of non-decomposable array response is the
uniform concentric circular array (UCCA). Denote $J$ and $L$
respectively as the number of rings and the number of antennas equally
placed on each ring in the array. As shown in \cite{VT02}, the array
response of UCCA can be expressed as
\begin{align}
\pmb{a}(\Theta_{n,m})=[\pmb{a}(\varphi_{1})^T,\cdots
 \pmb{a}(\varphi_{L})^T]^T\label{decopmi}
\end{align}
with\begin{align} \pmb{a}(\varphi_{l})\!&=\! [e^{j2\pi
\frac{d_1}{\lambda}{\cos(\phi_{n,m}-\varphi_{l})\cos\theta_{n,m}}},\nonumber\\
&\!\cdots\!, \!e^{j2\pi
\frac{d_{J}}{\lambda}{\cos(\phi_{n,m}-\varphi_{l})\cos\theta_{n,m}}}]^T\nonumber
\end{align}
where $d_j$ and $\varphi_{l}=2l\pi/L$
are respectively the radius of the $j$th ring and
the $l$th radial direction in the array, $j=1,\ldots,J$, $l=1,\ldots,L$, $\theta_{n,m}$ and
$\phi_{n,m}$ are respectively the directions of the $m$th ray in the $n$th
cluster with respect to the positive x- and
y-axis in the array.

The CDI of 3D MIMO channel is $\bar{\pmb{h}}= {\pmb{h}}/|\pmb{h}|$,
which is of unit-norm \cite{Love2003}. In limited feedback MIMO
systems, the CDI is quantized at the user by using a pre-determined
codebook and then fed back to the BS. The quality of CDI available
at the BS largely depends on the codebook. A desirable codebook
should be judiciously designed by taking the channel features into
account. To observe the unique features of 3D MIMO channels, besides
the vector expression $\pmb{h}\in \mathbb{C}^{N_t\times1}$, we also
express the channel in matrix form.

Taking the URA as an example, the 3D
MIMO channel and the array response can be expressed as
\begin{align}
\pmb{H}\in \mathbb{C}^{N_h\times N_v} ~\text{and}~ \pmb{A}_{n,m}=
\pmb{a}_{h}(\theta_{n,m})\pmb{a}_{v}(\phi_{n,m}) ^T\label{ary}
\end{align}
where $\pmb{h}=\text{vec}(\pmb{H})$,
$\pmb{a}(\Theta_{n,m})=\text{vec}(\pmb{A}_{n,m})$, and
$\text{vec}(\cdot)$ denotes the operation of vectorizing a matrix
into a vector. Then, the CDI can be expressed as $\bar{\pmb{H}}=\pmb{H}/\|\pmb{H}\|$.

The matrix representation helps us to identify the
different types of channel information in 3D MIMO channels with the
URA, since the columns and rows of $\pmb{H}$ respectively stand for the sub-responses
 at the horizontal and vertical directions.

If the ray responses of an array are decomposable, i.e.,
$\pmb{a}(\Theta_{n,m})=\pmb{x}\otimes\pmb{y}$, we can easily find
the matrix expression for each array response as $\pmb{x}\pmb{y}^T$
and the matrix expression for the channel, where
$\pmb{x}$ and $\pmb{y}$ are not necessarily the sub-responses in the
horizontal and vertical directions. However, the matrix expressions
for arbitrary antenna arrays can not be obtained straightforwardly and will be
discussed in Section IV.

\section{Codebook Design}
In this section,  we propose
a joint quantization codebook for spatially correlated 3D MIMO
channels, given that the channel matrix expression $\pmb{H}$ is available.

When tens or hundreds of antennas are
placed in the array, say $N_t=64$ or $256$, the dimension of vectors $\pmb{a}(\Theta_{n,m})$ and
$\pmb{h}$ can be very large, and the dimension of corresponding channel correlation matrix is much larger  since
\begin{align}
\pmb{R}=\pmb{E}\{\pmb{h}\pmb{h}^H\}\label{frmat}
\end{align}
which is of size $N_t\times N_t$. In fact, the dimension $N_t\times
N_t$ is large even for small value of $N_h$ and $N_v$, say $N_t=N_hN_v=8^2$ for  $N_h =N_v =8$. Such a  large dimension not only increases the computational
complexity for MIMO signal processing but also results in the difficultly of CDI
feedback \cite{Choi2013}.

The channel correlation matrix $\pmb{R}$ is important for many
codebooks \cite{Love2006,Ko2009,Choi2013a}. For example, when $\pmb{R}$ is
available, the rotated codebook  can immediately be
applied to quantize the 3D MIMO CDI vector, $\pmb{\bar h}$.
Specifically, the \emph{globally rotated codeword} can be constructed from an
instantaneous codeword $\pmb{g}_{0}$ of size $N_t\times1$ as
\cite{Love2006}
\begin{align}
\pmb{c}_g=\pmb{R}^{1/2}\pmb{g}_{0}\label{rote}
\end{align}
where the codeword $\pmb{c}_g$ is of unit-norm.

The globally rotated codebook has been proved to be asymptotical
optimal to quantize arbitrary spatially correlated channels
\cite{Rao2006}. Yet it is challenging to obtain the channel
correlation matrix $\pmb{R}$ at the BS since $N_t$ is large in 3D
MIMO systems. In the sequel, we strive to reduce the dimension of quantization by exploiting unique features of 3D MIMO channels.

\subsection{Different Types of CDI of 3D MIMO Channels}
We begin with the following proposition to identify different types of
CDI in 3D MIMO channels.
\newtheorem{proposition}{Proposition}
\begin{proposition} Any 3D MIMO channel matrix $\pmb{H}$ can be
decomposed into $\pmb{H}=\pmb{U}_h\pmb{H}_t\pmb{U}_v^T$ such that
$\pmb{E}\{\pmb{H}_t\pmb{H}_t^H\}=\text{diag}(\pmb{\lambda}_h)$ and
$\pmb{E}\{\pmb{H}_t^T\pmb{H}_t^\ast\}=\text{diag}(\pmb{\lambda}_v)$,
where $\pmb{U}_h$ and $\pmb{U}_v$ are unitary matrices,
$\pmb{\lambda}_h$ and $\pmb{\lambda}_v$ are vectors with nonnegative
entries.
\end{proposition}
\vspace{1mm} \textbf{Proof}: The proposition is easy to show by taking the SVD to channel matrix, which
is given here simply for introducing notations.

Denote the SVD of the left and right
correlation matrices of 3D MIMO channel respectively as
\begin{align}
\pmb{R}_h &= {\pmb{E}}\{\pmb{H}\pmb{H}^H\}=\pmb{U}_h\text{diag}(\pmb{\lambda}_h)\pmb{U}_h^H\label{lefcor}\\
\pmb{R}_v &= {\pmb{E}}\{\pmb{H}^T
\pmb{H}^\ast\}=\pmb{U}_v\text{diag}(\pmb{\lambda}_v)\pmb{U}_v^H\label{rigcor}
\end{align}
where the entries of $\pmb{\lambda}_h$ and $\pmb{\lambda}_v$ are in
a descending order. Considering that
$\pmb{H}_t=\pmb{U}_h^H\pmb{H}\pmb{U}_v^\ast$, and by using
\eqref{lefcor} and \eqref{rigcor}, we can obtain the proposition.
$\blacksquare$

This proposition suggests that \emph{for arbitrary 3D MIMO systems
under arbitrary channels}, we can always obtain two unitary matrices
shown in \eqref{lefcor} and  \eqref{rigcor}. Although this is
nothing but taking the SVD to the 3D MIMO channel matrix, but the
geometrical structure of the array has been explicitly exploited.

Note that we can also obtain a unitary matrix $\pmb{U}$ for the full
channel correlation matrix by the SVD as
$\pmb{R}=\pmb{U}\text{diag}(\pmb{\lambda})\pmb{U}^H$. To emphasize
the difference, the unitary matrix $\pmb{U}$ is referred to as
\emph{statistical direction information}, and the two unitary
matrices $\pmb{U}_h$, $\pmb{U}_v$ are referred to as
\emph{statistical sub-direction information}.

The statistical direction information can transform the 3D MIMO channel matrix
$\pmb{H}$ into at most $N_hN_v$ uncorrelated channel gains expressed in the
diagonal matrix $\text{diag}(\pmb{\lambda})$. In contrast, the two statistical sub-direction
information independently transform the channel matrix $\pmb{H}$ into at
most $N_h$ and $N_v$ uncorrelated channel gains respectively expressed in diagonal
matrices $\text{diag}(\pmb{\lambda}_h)$ and $\text{diag}(\pmb{\lambda}_v)$. One desirable property of statistical sub-direction information is
that the  dimensions of the correlation matrices of $N_h\times N_h$ and $N_v\times N_v$ are much smaller
than $N_hN_v\times N_hN_v$ for statistical direction
information. In the sequel, we refer to the two sub-directions as
``horizontal'' and ``vertical'' directions, although such notions only agree with their
physical meanings for the URA.

Denote $\pmb{\Lambda}_{i,j}$ as the the average channel gain of the
$(i,j)$th element in $\pmb{H}_{t}$ given by Proposition 1, e.g.,
$\pmb{\Lambda}^2_{i,j}\!=\!{\pmb{E}}\{|\pmb{H}_{t,i,j}|^2\}$. By
expressing $\pmb{u}_{i}$ and $\pmb{v}_{i}$ respectively as the $i$th
column of matrix $\pmb{U}_{h}$ and $\pmb{U}_{v}$, then
$\pmb{\Lambda}_{i,j}$ can be also expressed by
\begin{align}
\pmb{\Lambda}_{i,j}\!=\!{\pmb{E}}^{\frac{1}{2}}\{|\pmb{u}^H_i
\pmb{H}\pmb{v}^\ast_j|^2\}\label{fr}.
\end{align}
which means $\pmb{\Lambda}_{i,j}$ is the common average channel gain
shared by the $i$th statistical horizontal direction and $j$th
statistical vertical direction.

Different from $\pmb{\lambda}_h$ in \eqref{lefcor} and
$\pmb{\lambda}_v$ in \eqref{rigcor} which are obtained separately,
$\pmb{\Lambda}$ is interacted jointly with $\pmb{U}_h$ and
$\pmb{U}_v$, and can be interpreted as the statistical power
coupling information.

\subsection{The Proposed Codeword Structure}
Based on Proposition 1 and the observations in last subsection, we
propose a joint quantization codeword structure for quantizing the
CDI of 3D MIMO channel in matrix form,
$\bar{\pmb{H}}=\pmb{H}/\|\pmb{H}\|$, by separating the statistical
sub-directions, the average and instantaneous channel gains, which
is
\begin{align}
\pmb{C}_J = \hat{\pmb{U}}_h(\hat{\pmb{\Lambda}}\odot \pmb{G}) \hat{\pmb{U}}_v^T
\label{CJS}
\end{align}
where the codeword $\pmb{C}_J$ is normalized to have unit norm.  By using
$\text{vec}(ABC^T)=(C\otimes A)\text{vec}(B)$, the codeword
structure for quantizing the CDI of 3D MIMO
channel in vector form $\bar{\pmb{h}}$ can be expressed  as
\begin{align}
\pmb{c}_J =
(\hat{\pmb{U}}_v\otimes\hat{\pmb{U}}_h)\text{diag}(\hat{\pmb{\lambda}})\pmb{g}
\label{CJS2}
\end{align}
where $\pmb{c}_J= \text{vec}(\pmb{C}_J)$, $\hat{\pmb{\lambda}}=
\text{vec}(\hat{\pmb{\Lambda}})$ and $\pmb{g}= \text{vec}(\pmb{G})$.

The role of each part of the codeword is as follows,
\begin{itemize}
\item unitary matrix $\hat{\pmb{U}}_h$ is of size
$N_h\times r_h$, which targets at the \emph{statistical
sub-direction information} $\pmb{U}_h$ in horizontal direction
($1\leq r_h\leq N_h$),
\item unitary matrix $\hat{\pmb{U}}_v$ is of size $N_v\times r_v$, which targets at the \emph{
statistical sub-direction information} $\pmb{U}_v$ in vertical
direction ($1\leq r_v\leq N_v$),
\item nonnegative scaler matrix $\hat{\pmb{\Lambda}}$ is of size $r_h\times r_v$, which targets at
the \emph{statistical power coupling information} $\pmb{\Lambda}$
between the horizontal and vertical directions,
\item instantaneous codeword $\pmb{G}$ is of size $r_h\times r_v$, which
quantizes the \emph{instantaneous power coupling information}. In
fact, the instantaneous power coupling information reflects the
instantaneous CDI of $\pmb{H}_t$ in Proposition 1.
\end{itemize}

With the codeword structure in \eqref{CJS}, the CDIs in horizontal
and vertical directions are jointly quantized together with the
power coupling. Therefore, we refer the
codebook of codewords with this structure as the \emph{joint CDI quantization codebook}.

Compared with the codeword structure of the globally rotated codebook shown in \eqref{rote}, the proposed codeword structure in \eqref{CJS2} needs low-dimensional statistical channel information for rotation.

In practice, the statistical information $\pmb{U}_h$, $\pmb{U}_v$
and $\pmb{\Lambda}$ can be obtained at the BS either by uplink
channel estimation or by feedback.

In FDD systems where the downlink and the uplink are operated at
separated frequency bands, estimating the downlink channel
correlation matrix from uplink training symbols is still possible
\cite{Maretta2001}. However, such a method may not be used to
estimate the channel correlation matrix $\pmb{R}$ for 3D MIMO
systems, which is of high dimension. The problem of estimating the
correlation matrix for high-dimension random vectors is recognized
as ``curse of dimensionality'' and far from trivial
\cite{Marzetta2011a}. This is because the dimension of the channel
correlation matrix may be comparable with the number of collected
training symbols such that the widely-used ``sample covariance''
estimation becomes invalid.

The matrices $\pmb{U}_h$, $\pmb{U}_v$ and
$\pmb{\Lambda}$ can also be quantized and fed back
\cite{Ghosh2012,Kris2014}. Since the spatial correlation information
can be fed back in  wide-band and in long-term, the feedback
overhead can be almost ignored compared with the feedback for
instantaneous CDI. This is especially true when the array feature is taken into account. For example, for the URA, according to Szego's theory of Toeplitz
matrices, the statistical sub-direction information  $\pmb{U}_h$
and $\pmb{U}_v$ become a subset of DFT matrix when the size of the array
increases \cite{Adhikary2013}. This will significantly simplify
the feedback for the correlation information.

Since our focus  is to reduce the dimension of quantization by introducing new codeword structure, we assume that
the statistical channel information ${\pmb{U}}_h$, ${\pmb{U}}_v$ and
${\pmb{\Lambda}}$ are available at the BS. Then, the matrices used to construct the codewod  with structure in \eqref{CJS}
$\hat{\pmb{U}}_h$, $\hat{\pmb{U}}_v$ and $\hat{\pmb{\Lambda}}$ can
be obtained as
\begin{align}
\hat{\pmb{U}}_h&=[\pmb{u}_{1},\cdots,\pmb{u}_{r_h}], \quad
\hat{\pmb{U}}_v=[\pmb{v}_{1},\cdots,\pmb{v}_{r_v}]\nonumber\\
\hat{\pmb{\Lambda}}_{i,j}&= \pmb{\Lambda}_{i,j}, \text{with~} 1\leq
i\leq r_h, 1\leq j\leq r_v \label{midcor}
\end{align}

The codeword quantizing the instantaneous CDI of 3D MIMO channel,
i.e., $\pmb{G}$, can be obtained by using the codebooks designed for
uncorrelated channels with dimension of $r_h\times r_v$. For
example, GSP codebooks proposed in \cite{Love2003} can be used,
which is optimized for uncorrelated Rayleigh channels.

The dimensions $r_h$ and $r_v$  in  \eqref{CJS} can be designed to
further reduce the dimension of codeword by discarding trivial
statistical sub-directions in strong correlated channel. We leave
this topic for future study, since similar idea has already been
developed in \cite{Ko2009}, and moreover, the dimension parameter
can be optimized by considering various system parameters, e.g., the
error tolerance, storage space, and overall quantization
performance.

\subsection{Extension to Multi-antenna Users}
When considering multiple receive antennas at the user, it is reasonable to assume that the statistical channel information in \eqref{CJS} associated with
different receive antennas are identical
 due to the small space separation. However, the
instantaneous CDI associated with different receive antennas may
differ.
Based on this observation, we can easily extend the codebook design
to the systems where each user is equipped with $N_r$ antennas.

Denote the super-script in parentheses as the index of receiver
antenna, the joint quantization codeword in \eqref{CJS2} for
multi-antenna users can be given as
\begin{align}
\pmb{c}_J^{(1,\dots, N_r)}\!&=\![\pmb{c}_J^{(1)},\!\cdots\!,\!\pmb{c}_J^{(N_r)}]\nonumber\\
&\!=(\hat{\pmb{U}}_v\otimes\hat{\pmb{U}}_h){\text{diag}(\hat{\pmb{\lambda}})}[
\pmb{g}^{(1)},\!\cdots\!,\!\pmb{g}^{(N_r)}]\nonumber
\end{align}
where $\pmb{g}^{(1)},\cdots,\pmb{g}^{(N_r)}$ are respectively the
instantaneous codewords for different receive antennas.

%

\section{Statistical Information for Arbitrary Antenna Array}
The proposed codeword structure is applicable for any antenna array if the
channel can be expressed in matrix form as $\pmb{H}$. This is because Proposition 1 is valid for arbitrary 3D MIMO channels. Although quite natural to the
decomposable  arrays such as the URA, the way to express the channels in matrix form
for arbitrary antenna array is not obvious because of
two  issues.

The first issue is the choice of dimensionality for the 3D MIMO
channel matrix, i.e., $N_h$ and $N_v$ with $N_h\times N_v=N_t$.
Taking the UCCA array with $N_t=32$ as an example, we may express
the channel vector $\pmb{h}$ in  a matrix form $\pmb{H}$ of size
$8\times4$, $4\times8$, or the others, but we are not clear which
expression achieves a better quantization performance because  the
array is not rectangular.

The second issue is the ``antenna grouping'' in the channel matrix,
i.e., which subset of antenna responses should be in the same
row or column in $\pmb{H}$ for a given dimensionality of $N_h$ and
$N_v$. The antenna grouping determines the values of
$\pmb{U}_h$, $\pmb{U}_v$ and $\pmb{\Lambda}$, and eventually affects the
performance of the proposed codebook.

As a consequence, we need to solve these two issues when designing
the joint codebook for 3D MIMO systems with arbitrary antenna array.
The challenges for addressing the two issues are different. The
choices satisfying $N_t=N_h\times N_v$ are always limited, and thus
exhaustive searching is efficient to find the best dimensionality.
However, the possible choices of antenna grouping exponentially
increases with $N_t$, which belongs to a combinational problem and
is of prohibitive computational complexity. Therefore, we focus on
the second issue in the following, i.e., to find proper $\pmb{U}_h$,
$\pmb{U}_v$ and $\pmb{\Lambda}$ for a given $N_h$ and $N_v$.

\subsection{Optimizing $\pmb{U}_h$, $\pmb{U}_v$ and
$\pmb{\Lambda}$} To find the desirable statistical information,
before solving the antenna grouping problem, we first reconsider the
proposed codebook structure. Recall that the entries of $\pmb{H}_t$
defined in Proposition 1 reflect the instantaneous channel gains,
and  the GSP codebook can be used to quantize the instantaneous CDI,
which is optimal for i.i.d. channels. Therefore, the proposed
codebook will be optimal if the entries of $\pmb{H}_t$ are
uncorrelated and the statistical information is perfect (i.e.,
$\hat{\pmb{U}}_h=\pmb{U}_{h}$, $\hat{\pmb{U}}_h= \pmb{U}_{v}$ and
$\hat{\pmb{\Lambda}}=\pmb{\Lambda}$).

We can reconstruct a full channel matrix $\hat{\pmb{R}}$  from the
matrices with lower dimensions as
\begin{align}
\hat{\pmb{R}}&=\pmb{E}\{\text{vec}(\pmb{U}_h\pmb{H}_t\pmb{U}_v^T)\text{vec}(
\pmb{U}_h\pmb{H}_t\pmb{U}_v^T)^H\}\nonumber\\
&=(\pmb{U}_v\otimes\pmb{U}_h)\pmb{E}\{\text{vec}\{\pmb{H}_t\}\text{vec}\{\pmb{H}_t\}^H\}
(\pmb{U}_v\otimes\pmb{U}_h)^H\nonumber\\
&=(\pmb{U}_v\!\otimes\!\pmb{U}_h)\text{diag}(\text{vec}\{\pmb{\Lambda}\!\odot\!\pmb{\Lambda}\})
(\pmb{U}_v\!\otimes\!\pmb{U}_h)^H \label{rcr}
\end{align}
where \eqref{rcr} is achieved when the entries of $\pmb{H}_t$ are
uncorrelated.

If $\hat{\pmb{R}}=\pmb{R}$, the proposed codebook is identical to
the globally rotated codebook, since the codeword in \eqref{CJS2}
becomes $\pmb{c}_J=\hat{\pmb{R}}^{\frac{1}{2}}\pmb{g}$ and the same
as in \eqref{rote}. Unfortunately, in general cases,
$\hat{\pmb{R}}\neq\pmb{R}$, and the proposed codebook becomes
inferior  to the globally rotated codebook, which is asymptotically
optimal. To provide good quantization performance, it is reasonable
to find $\pmb{U}_h$, $\pmb{U}_v$ and $\pmb{\Lambda}$ such that
$\hat{\pmb{R}}$ is as close to $\pmb{R}$ as possible.

For simplicity, we define
$\pmb{\lambda}_t=\text{vec}\{\pmb{\Lambda}\odot\pmb{\Lambda}\}$. The
problem of finding the optimal $\pmb{U}_h$, $\pmb{U}_v$ and
$\pmb{\lambda}_t$ for arbitrary antenna array can be modeled as
\begin{align}
\min_{\pmb{U}_h,\pmb{U}_v,\pmb{\lambda}_t}
&\|\pmb{R}\!-\!(\pmb{U}_v\!\otimes\!\pmb{U}_h)\text{diag}(\pmb{\lambda}_t)
(\pmb{U}_v\!\otimes\!\pmb{U}_h)^H\|^2_F\label{kp}\\
\text{s.t.}& ~~\quad \pmb{U}_v^H\pmb{U}_v=\pmb{I}_{N_v},\nonumber\\
&~~ \quad\pmb{U}_h^H\pmb{U}_h=\pmb{I}_{N_h},\nonumber\\
&~~   \quad\pmb{\lambda}_t\succ0\nonumber
\end{align}
where $\pmb{I}_{n}$ is the identity matrix of size $n\times n$, and
$\pmb{x}\succ 0$ means each element in $\pmb{x}$ is larger than $0$.
The problem in \eqref{kp} does not directly optimize
the antenna grouping in the channel matrix, which is unnecessary since when given the
optimal solution of $\pmb{U}_h$,  $\pmb{U}_v$ and
$\pmb{\lambda}_t$, we
can obtain the matrices used to construct the joint quantization codewods.

Problem \eqref{kp} belongs to a classic approximation problem
of Tucker decomposition \cite{Stefan2012}, where $\pmb{\lambda}_t$
is the core tensor, $\pmb{U}_h$ and $\pmb{U}_v$ are respectively
the factor matrices with columns $\pmb{u}_i$ and $\pmb{v}_i$ as tensors.
The main objective of Tucker decomposition is to decompose a higher
dimensional matrix into low dimensional factor matrices, and the
tensor core encompass all the possible interactions among the low
dimensional tensors in the factor matrices. In other words, Tucker
decomposition is  to reduce the dimension in the large matrix by finding the structure
properties. Moreover, Tucker decomposition is a
generation of the matrix SVD with several desirable
features, such as orthogonality, decorrelation and
computational tractability. However, in general the problem of finding the
solution of Tucker decomposition is NP-hard,
and there are few efficient algorithms in use.

Note that although the problem in \eqref{kp}  contains
three variables, it can be simplified by only
optimizing $\pmb{U}_h$ and $\pmb{U}_v$  without losing the optimality.
This is because given $\pmb{U}_h$ and $\pmb{U}_v$, the objective
function satisfies
\begin{align}
&\quad\min_{\pmb{\lambda}_t}
\|\pmb{R}\!-\!(\pmb{U}_v\!\otimes\!\pmb{U}_h)\text{diag}(\pmb{\lambda}_t)(\pmb{U}_v\!\otimes\!\pmb{U}_h)^H\|_F^2\nonumber
\\&=\min_{\pmb{\lambda}_t}\!\|(\pmb{U}_v\!\otimes\pmb{U}_h)^H\pmb{R}(\pmb{U}_v\!\otimes\!\pmb{U}_h)\!\!-\!\text{diag}(\pmb{\lambda}_t)\|_F^2\label{eq2}
\\&=\min_{\pmb{\lambda}_t}\!
\|\text{diag}\left((\pmb{U}_v\!\otimes\pmb{U}_h)^H\pmb{R}(\pmb{U}_v\!\otimes\!\pmb{U}_h)\right)\!\!-\!\text{diag}(\pmb{\lambda}_t)\|_F^2\nonumber\\
&\quad+\|\text{off}\left((\pmb{U}_v\otimes\pmb{U}_h)^H\pmb{R}(\pmb{U}_v\otimes\pmb{U}_h)\right)\|_F^2\nonumber
\\&= \|\text{off}\left((\pmb{U}_v\otimes\pmb{U}_h)^H\pmb{R}(\pmb{U}_v\otimes\pmb{U}_h)\right)\|_F^2\label{eq3}
\end{align}
where \eqref{eq2} is due to fact that the norm is unitarily
invariant, $\text{off}(\pmb{X})$ is the operation to the matrix
$\pmb{X}$ with all zeros on the diagonal, \eqref{eq3} is achieved by
the optimal $\pmb{\lambda}_t$ for a given $\pmb{U}_h$ and
$\pmb{U}_v$, which is
\begin{align}
\text{diag}(\pmb{\lambda}_t) =
\text{diag}\left((\pmb{U}_v\otimes\pmb{U}_h)^H\pmb{R}(\pmb{U}_v\otimes\pmb{U}_h)\right)\label{lmdx}
\end{align}
It can be verified that the operation for computing $\pmb{\lambda}_t$
in \eqref{lmdx} is identical to that for computing $\pmb{\Lambda}$
with \eqref{fr}.

Since \eqref{eq3} does not depend on the
parameter $\pmb{\lambda}_t$, we only need to find
$\pmb{U}_h$ and $\pmb{U}_v$ from a new optimization problem by replacing the objective function in \eqref{kp} with \eqref{eq3}. With the optimized $\pmb{U}_h$ and $\pmb{U}_v$, we can immediately obtaining optimal
$\pmb{\lambda}_t$ by using \eqref{lmdx}.

\subsection{Closed-form Solution for $\pmb{U}_h$ and $\pmb{U}_v$}
Since a closed-form solution is more desirable for practical use, we
consider a modified problem to find $\pmb{U}_h$ and $\pmb{U}_v$.
Specifically, we impose an extra constraint on $\pmb{\lambda}_t$
into the new optimization problem, which is given by
\begin{align}
\pmb{\lambda}_t = \pmb{\lambda}_v\otimes\pmb{\lambda}_h\label{acc}
\end{align}
where $\pmb{\lambda}_v\succ0$ and $\pmb{\lambda}_h\succ0$. Then, the
term inside the objective function of problem \eqref{kp} becomes
\begin{align}
\!(\pmb{U}_v\!\otimes\!\pmb{U}_h)\text{diag}(\pmb{\lambda}_t)
(\pmb{U}_v\!\otimes\!\pmb{U}_h)^H=\pmb{B}\otimes\pmb{C}\label{eq19}
\end{align}
where $\pmb{B}=\pmb{U}_v\text{diag}(\pmb{\lambda}_v)\pmb{U}_v^H$ and
$\pmb{C}=\pmb{U}_h\text{diag}(\pmb{\lambda}_h)\pmb{U}_h^H$ are
positive semi-definite matrices.

With the constraint in \eqref{acc} and considering \eqref{eq19}, the optimization problem to find $\pmb{U}_h$ and $\pmb{U}_v$ becomes
\begin{align}
\min_{B,C} &~~ \|\pmb{R}-\pmb{B} \otimes
\pmb{C}\|^2_F\label{kp6}\\
\text{s.t.}& ~~\pmb{B}\in \mathbb{S}^{N_v\times N_v}, \text{and}~
\pmb{C}\in \mathbb{S}^{N_h\times N_h}\nonumber
\end{align}
where $\mathbb{S}^{n\times n}$ is the space of positive
semi-definite matrices with the dimensionality of $n\times n$.

The new problem in \eqref{kp6} offers a suboptimal solution of
$\pmb{U}_h$ and $\pmb{U}_v$ with closed-form for the Tucker
decomposition, and is known as the Kronecker product decomposition
\cite{Nikos1997}. There are many other benefits to consider the
Kronecker product decomposition here. By using such an
decomposition, many structure properties of $\pmb{R}$, such as
symmetry, definite, and permutations, can be inherited by the
matrices $\pmb{B}$ and $\pmb{C}$. It is worthy to note that these
structure properties are usually led by the regularity of antenna
array, e.g., symmetry, and nested subarrays.

The solution to the problem in \eqref{kp6} is given in
\cite{Nikos1997}. To obtain the solution, according to
\cite{Nikos1997}, we need to rearrange the matrix $\pmb{R}$.
Specifically, the matrix $\pmb{R}$ is divided into $N_v\times N_v$
blocks as
\begin{align}
  \pmb{R} = \left(
              \begin{array}{ccc}
                \pmb{R}_{1,1} & \cdots & \pmb{R}_{1,N_v} \\
                \vdots & \ddots & \vdots \\
                \pmb{R}_{N_v,1} & \cdots & \pmb{R}_{N_v,N_v} \\
              \end{array}
            \right)
\end{align}
where the $(i,j)$th block denoted as $\pmb{R}_{i,j}$ is of size
$N_h\times N_h$. Then a rearranged matrix is generated by
\begin{align}
  \tilde{\pmb{R}}\!=[\text{vec}(\pmb{R}_{1,1}),\text{vec}(\pmb{R}_{2,1}),\!\cdots,\!\text{vec}(\pmb{R}_{N_v,N_v})]^T
\end{align}
which is of size $N_v^2\times N_h^2$.

Denote the largest singular value, the corresponding left and right
eigenvectors of the SVD to $\tilde{\pmb{R}}$ respectively as
$\sigma^2$, $\pmb{u}$ and $\pmb{v}$. Then, as shown in
\cite{Nikos1997}, the matrices $\pmb{B}$ and $\pmb{C}$ are obtained
as
\begin{align}
\text{vec}(\pmb{B})=\sigma\pmb{u}, \text{and~}
\text{vec}(\pmb{C})=\sigma\pmb{v}
\end{align}
Moreover, Since $\pmb{R}$ is symmetric and positive semi-definite,
$\pmb{B}$ and $\pmb{C}$ are also symmetric and positive
semi-definite\cite{Nikos1997}. Thus, the matrices $\pmb{B}$ and
$\pmb{C}$ can be regarded as the correlation matrices.

By letting $\pmb{R}_v=\pmb{B}$, $\pmb{R}_h=\pmb{C}$, using the SVD,
we can obtain $\pmb{U}_v$ and $\pmb{U}_h$ immediately. Then, by using
\eqref{lmdx}, we can obtain the optimal
$\pmb{\lambda}_t$ or $\pmb{\Lambda}$.

It is easy to validate that the statistical direction information
$\pmb{U}$ is approximated by the lower dimension matrices as $\pmb{U}_v
\otimes \pmb{U}_h$, and the average channel gains in $\pmb{\lambda}$
are approximated by $\pmb{\lambda}_t$. By introducing the structure of joint
quantization codebook, we employ a new correlation matrix constructed with reduced
dimension that approximates the full correlation matrix as close as
possible. Therefore, with the obtained correlation matrices, we can
not only solve the dimension problem for rotated codebook in
\cite{Love2006}, but also for the codebook in \cite{Choi2013a}, both rely on the channel correlation matrices.

\section{Performance Analysis}
Since the globally rotated codebook is asymptotical optimal and can
serve as a performance upper bound for other codebooks, we analyze the performance of the proposed joint CDI quantization codebook by comparing
with the globally rotated codebook in this section.

As addressed in Section IV.A, if the entries of $\pmb{H}_t$
are uncorrelated and the statistical information used in the codeword with structure in \eqref{CJS} is
perfect, the joint CDI quantization codebook will perform the same as the globally rotated codebook.
However, it is unclear whether the entries of
$\pmb{H}_t$ are uncorrelated or not. In what follows, we show that for 3D MIMO systems with URA such an
uncorrelated property is valid
under very general channel conditions.

To facilitate analysis and gain useful insights, we consider weak
i.i.d. rays in 3D MIMO channels, which assume,
\begin{enumerate}
\item the angles $\theta_{n,m}$ are i.i.d. for $\forall$ $n$ and
$m$,
\item the angles $\phi_{n,m}$ are i.i.d. for $\forall$ $n$ and
$m$,
\item the gains $g_{n,m}$ are uncorrelated for $\forall$ $n$ and
$m$,
\end{enumerate}
where the word ``weak'' comes from the fact that the third condition only requires the gains being uncorrelated but not being identically distributed.

The above assumption does not weaken our performance analysis for
realistic scenarios. First of all, the assumption of uncorrelated
ray gains follows from the commonly-used uncorrelated scattering
assumption for multi-path fading channels, which is validated
by many channel models \cite{Yong2003, WINNER2,rel12cm}. Second,
i.i.d. angles are observed very typical and thus are adopted in the 3GPP
channel models \cite{rel12cm}. Finally, no specific probability density
distribution assumption is imposed on the gains and angles. For example,
the angles can be uniformly distributed \cite{Yong2003},
Gaussian distributed \cite{WINNER2}, or log-normal distributed
\cite{rel12cm}.

In the following, we show that the proposed joint CDI quantization
codebook achieves the same performance as the globally rotated
codebook when the 3D MIMO channels have weak i.i.d. rays and the BS
is equipped with URA.

\vspace{2mm}
\newtheorem{lemma}{Lemma}
\begin{lemma} For a 3D MIMO system with URA, the channel with weak i.i.d.
rays can be expressed as $\pmb{H}=\acute{\pmb{H}}\pmb{U}_v^T$,
where $\pmb{U}_v$ is a unitary matrix and the columns of
$\acute{\pmb{H}}$ are uncorrelated with each other.\end{lemma}

\vspace{1mm} \textbf{Proof}: See Appendix \ref{Proof Lemma1}. $\blacksquare$

\vspace{2mm} \begin{lemma} For a 3D MIMO system with URA, the
channel with weak i.i.d. rays can be expressed as
$\pmb{H}=\pmb{U}_h\grave{\pmb{H}}$,  where $\pmb{U}_h$ is a unitary
matrix and the rows of $\grave{\pmb{H}}$ are uncorrelated with each
other.\end{lemma}

\textbf{Proof}: The proof is similar to Lemma $1$.$\blacksquare$

\vspace{2mm}
\begin{lemma} For a 3D MIMO system with URA under the
channel with weak i.i.d. rays, the entries of $\pmb{H}_t$ in the channel decomposition $\pmb{H}=\pmb{U}_h\pmb{H}_t\pmb{U}_v^T$
given by Proposition 1 are uncorrelated.
\end{lemma}

\textbf{Proof}: See Appendix \ref{Proof Lemma3}.  $\blacksquare$

\vspace{2mm} By using these lemmas for the 3D MIMO system with URA
under channels with weak i.i.d. rays, we can immediately find the
relationship between the joint CDI quantization codebook and the
globally rotated codebook.

\newtheorem{theorem}{Theorem}
\vspace{2mm}
\begin{theorem} For a 3D MIMO system with URA under the channel   with weak i.i.d. rays, the proposed joint
quantization codebook with perfect correlation matrices performs the same as the globally rotated codebook.
\end{theorem}

\textbf{Proof}: See Appendix \ref{Proof Theorem}.$\blacksquare$

\vspace{2mm} It follows that  for 3D MIMO systems with URA under the
channels with i.i.d. rays, the proposed joint CDI quantization
codebook is asymptotically optimal as the codebook size increases,
which is promised by the globally rotated codebook \cite{Rao2006}.

It is worthy to note that the Theorem 1 provides a sufficient
condition on the 3D MIMO channels where the joint CDI quantization
and globally rotated codebook are identical. The entries in
$\pmb{H}_t$ may not be uncorrelated when other antenna arrays ( for
example, UCCA) are used. This means for general conditions, the two
codebooks are not identical. In these cases, the joint CDI
quantization codebook, though still applicable, becomes inferior to
the globally rotated codebook.

\subsection{Further Comparison With Globally Rotated
Codebook}
 As shown in \eqref{frmat}, the globally rotated
codebook employs the channel correlation matrix, which is of size
$N_hN_v\times N_hN_v$. By contrast, the joint CDI quantization
codebook employs three lower dimension correlation matrices, each
respectively of size $N_h\times N_h$ for $\pmb{R}_h$ given by
\eqref{lefcor}, of size $N_v\times N_v$ for $\pmb{R}_v$ given by
\eqref{rigcor}, and of size $N_h\times N_v$ for $\pmb{\Lambda}$
given by \eqref{midcor}. Apparently, a high dimension matrix leads
to a high complexity in generating the codewords to quantize the
statistical information or transforming the instantaneous codeword.
By using the joint CDI quantization, the computational complexity is
significantly reduced.

In addition, different size of correlation matrix needs different
amount of  information to estimate or feedback. Taking an $8\times8$
3D MIMO channel as an example, the globally rotated codebook
requires a correlation matrix with $64^2=4096$ entries, while the
joint CDI quantization requires the correlation matrix only with
$64\times3=192$ entries. Even when the correlation matrices
$\pmb{R}$, $\pmb{R}_h$, $\pmb{R}_v$ are Hermitian and Toeplitz, the
globally rotated codebook needs $63$ complex elements in $\pmb{R}$,
while the joint quantization only needs $7$ complex elements in
$\pmb{R}_h$, $7$ in $\pmb{R}_v$, and $r_h\times r_v$ scaler elements
in $\pmb{\Lambda}$. This implies that the proposed joint CDI
quantization book requires much less channel correlation information
to be estimated or fed back than the globally rotated codebook.

\subsection{Comparison With Independent CDI
Quantization} Another strategy, independent CDI quantization, can
be applied to 3D MIMO systems, where the CDIs in horizontal and
vertical directions are quantized independently \cite{ALSB2011}. By
nature, joint quantization is not a simple aggregation of
independent quantization in two directions.

Denote the codewords independently designed for horizontal and
vertical directions as  $\pmb{c}_{Ih}$ and $\pmb{c}_{Iv}$,
respectively. In \cite{ALSB2011}, the CDI of 3D MIMO is
reconstructed by two independently designed codewords as
$\pmb{C}_I=\pmb{c}_{Ih}\pmb{c}_{Iv}^T$. Taking independently rotated
codebook as an example, $\pmb{c}_{Ih}$ and  $\pmb{c}_{Iv}$ are
respectively obtained by normalizing $\pmb{R}_h^{1/2}\pmb{c}_{h}$
and $ \pmb{R}_v^{1/2}\pmb{c}_{v}$, where $\pmb{c}_{h}$ and
$\pmb{c}_{v}$ are instantaneous codewords for horizontal and
vertical directions. The independent CDI quantization is easy to
implement and is compatible to existing 2D MIMO systems, but at a
cost of quantization performance loss.

First, independent CDI quantization will not be optimal to quantize
the 3D MIMO channel if the rank of channel matrix $\pmb{H}$ exceeds
one. The rank of $\pmb{H}$, denoted as $\text{rank}(\pmb{H})$, has a
close relationship with the scattering environment. If there are
multiple well-separated clusters or rays in the channel, it is easy
to see that $\text{rank}(\pmb{H})$ is far larger than one. However,
the independent CDI quantization codebook always yields
$\text{rank}(\pmb{C}_I)=1$, which fails to match the rank of generic
3D MIMO channel matrix. This implies that independent CDI
quantization is only appropriate for quantizing the rank  one 3D
MIMO channels with a single ray. By contrast, the row rank and
column rank of the joint CDI quantization codeword $\pmb{C}_J$  are
respectively $r_h$ and $r_v$. By selecting proper values of $r_h$
and $r_v$, we can design joint CDI quantization codeword for any 3D
MIMO channel matrix $\pmb{H}$. In practice, the values of $r_h$ and
$r_v$ may be selected less than the rank of $\pmb{H}$ for the
purpose of dimension reduction. In these cases, the performance loss
can be minimized by reserving the statistical directions in
\eqref{CJS} with nontrivial average channel gains.

Second, independent CDI quantization results in redundant
quantization on the channel information, since the codebooks are
separately designed for horizontal and vertical directions from
their own perspective. To see this, we can find that the power
profile of uncorrelated clusters seen at horizontal direction and at
vertical direction in the transformed 3D MIMO channel after SVD are
respectively the eigenvalues of $\pmb{R}_h$ and $\pmb{R}_v$ given by
\eqref{lefcor} and \eqref{rigcor}. The two power profiles are
generally not independent with each other. However, with independent
CDI quantization the two power profiles are quantized separately,
which leads to a redundancy in the CDI quantization. By contrast,
when using joint CDI quantization, such a redundancy can be
completely removed by quantizing the power coupling information. As
a consequence, joint CDI quantization is more efficient to quantize
the 3D MIMO channels than independent CDI quantization.

\section{Simulation Results}

In this section, we compare the performance of joint CDI
quantization with existing codebooks for 3D MIMO systems with planar array by simulations.

We consider multi-user MIMO transmission, where $K$ single-antenna
users are served by the BS at the same time-frequency resource with
zero-forcing beamforming \cite{WE08}. In the simulation, the users
are randomly selected, which is equivalent to be selected by
Round-Robin scheduling. The users are with homogeneous SNR but with different azimuth and elevation angles. In
order to put the results of different antenna array size within one
figure, the $X$-axis is set as the receive SNR.  For transmit SNR, an extra
$\log_2N_t$ dB should be considered due to the array gain.
The gains of each ray are i.i.d. complex Gaussian distributed with
zero mean. The sum rates are computed by the Shannon formula by
averaging over $10^3$ channel realizations.

In the following, we consider two different angular
spread modelings for the channel  given
by \eqref{CNM}. One
is simplified, which allows us to run simulations for large scale
system efficiently. The other is more realistic, which is given in
\cite{rel12cm} and allows us to obtain more reliable results. It should
be noted that our conclusions hold for both channel models.

In the simulations, all the instantaneous CDI codewords are
generated using random vector quantization (RVQ), which is easy to
generate while with performance close to the GSP codebook
\cite{Jindal2007}. Specifically, two $B$-bit RVQ codebooks are used
for quantizing the instantaneous horizonal and vertical channel
directions in the independently rotated codebook. A $2B$-bit RVQ
codebook are used for quantizing the instantaneous channel direction
information respectively in the joint CDI quantization codebooks,
and in the globally rotated codebook.

\subsection{Results in Simplified Scenarios}
In the simplified angular spread model, the azimuth and elevation angles
in  \eqref{CNM} are,
\begin{align}
\theta_{n,m}&= \theta_{0}+\theta_{n}+\delta \theta_{n,m}\\
\phi_{n,m}&=\phi_{0}+\phi_{n}+\delta \phi_{n,m}
\end{align}
where $\theta_{0}=\pmb{U}(-60^\circ,~60^\circ)$ and
$\phi_{0}=\pmb{U}(-45^\circ,~45^\circ)$ are respectively the mean of
azimuth and elevation clusters, and
$\pmb{U}(a,b)$ denotes uniform distribution with range from $a$ to
$b$,
$\theta_{n}$ and $\phi_{n}$ are the the azimuth and elevation
deviation of the $n$th cluster from their mean, $\delta\theta_{n,m}$
and $\delta\phi_{n,m}$ are the random offsets. We set $N=12$ and
$M=20$, and model the $\theta_{n}$ and $\phi_{n}$ as i.i.d.
Gaussian distributed with zero mean and a variance of $\sigma^2$,
where $\sigma$ is a parameter representing the angular spread. We model
$\delta\theta_{n,m}$ and $\delta\phi_{n,m}$ as Lapalacian
distributed with a root mean square of $1^{\circ}$. As shown by
\cite{WINNER2}, such a simplified channel model well preserves the spatial features of the channels modelled in
\cite{rel12cm}.

Both low angular spread ($\sigma=5^{\circ}$) and large angular spread
($\sigma=20^{\circ}$) scenarios are evaluated.

We evaluate the average sum rate of a 3D MIMO system with the
configuration $N_h$ and $N_v$ using the joint CDI quantization
codebook (JQC). For comparison, the performance of globally rotated
codebook (GRC) is provided, which quantizes a $N_t = N_h \times N_v$
3D MIMO CDI vector as a whole using the full correlation matrix
$\pmb{R}$ in \eqref{rote}. The performance of independent
quantization with rotated codebook (IQC) is given, which quantizes
an $N_h \times1$ CDI vector for each direction individually using
the statistical information $\pmb{R}_h$ in \eqref{lefcor} and
$\pmb{R}_v$ in \eqref{rigcor}.

\subsubsection{Performance for 3D MIMO Systems with URA}
We first evaluate the performance of joint CDI quantization codebook
for 3D MIMO system using the URA, where $N_h=N_v$.

\begin{figure}[t]
\centering
   \includegraphics[width=3.8in]{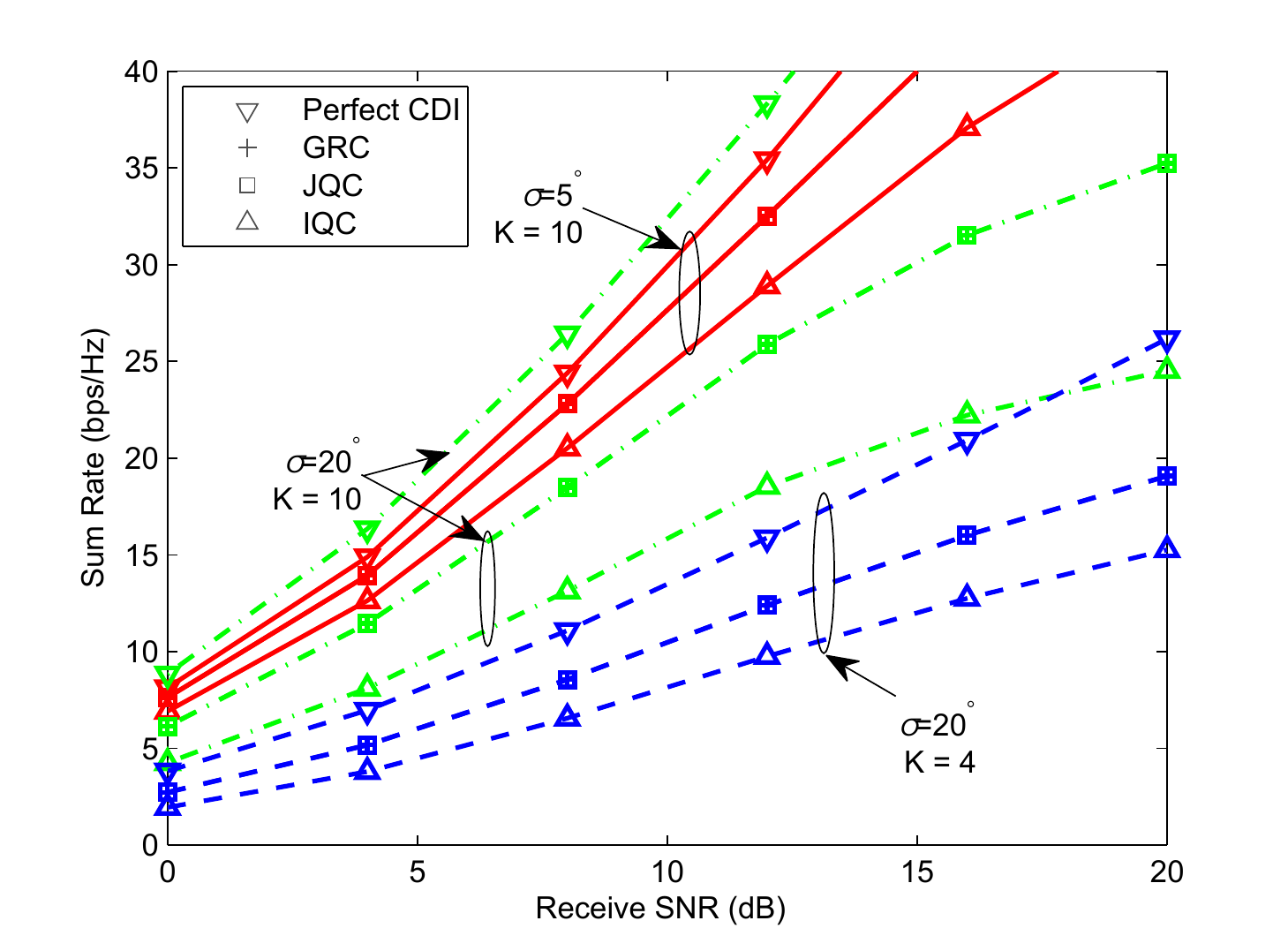}
   \caption{Average sum rates of 3D MIMO systems using URA for different angular spreads and numbers of users, where $N_h=N_v=8$ and $B=8$.}\label{KandAS}
\end{figure}

The average sum rates of the 3D MIMO systems using
different codebooks are shown in Fig. \ref{KandAS}, where all the channel statistical information are perfect. It is shown that
under different scenarios, the curves for the joint CDI
quantization codebook overlap with those of the globally
rotated codebook, which validates Theorem $1$. Under the same angular
spread, when the number of users grows, the
performance gap between the joint and independent CDI quantization
increases. Compared with the low angular spread, the performance gap
is larger in the channel with
large angular spread.

\begin{figure}[t]
\centering
   \includegraphics[width=3.8in]{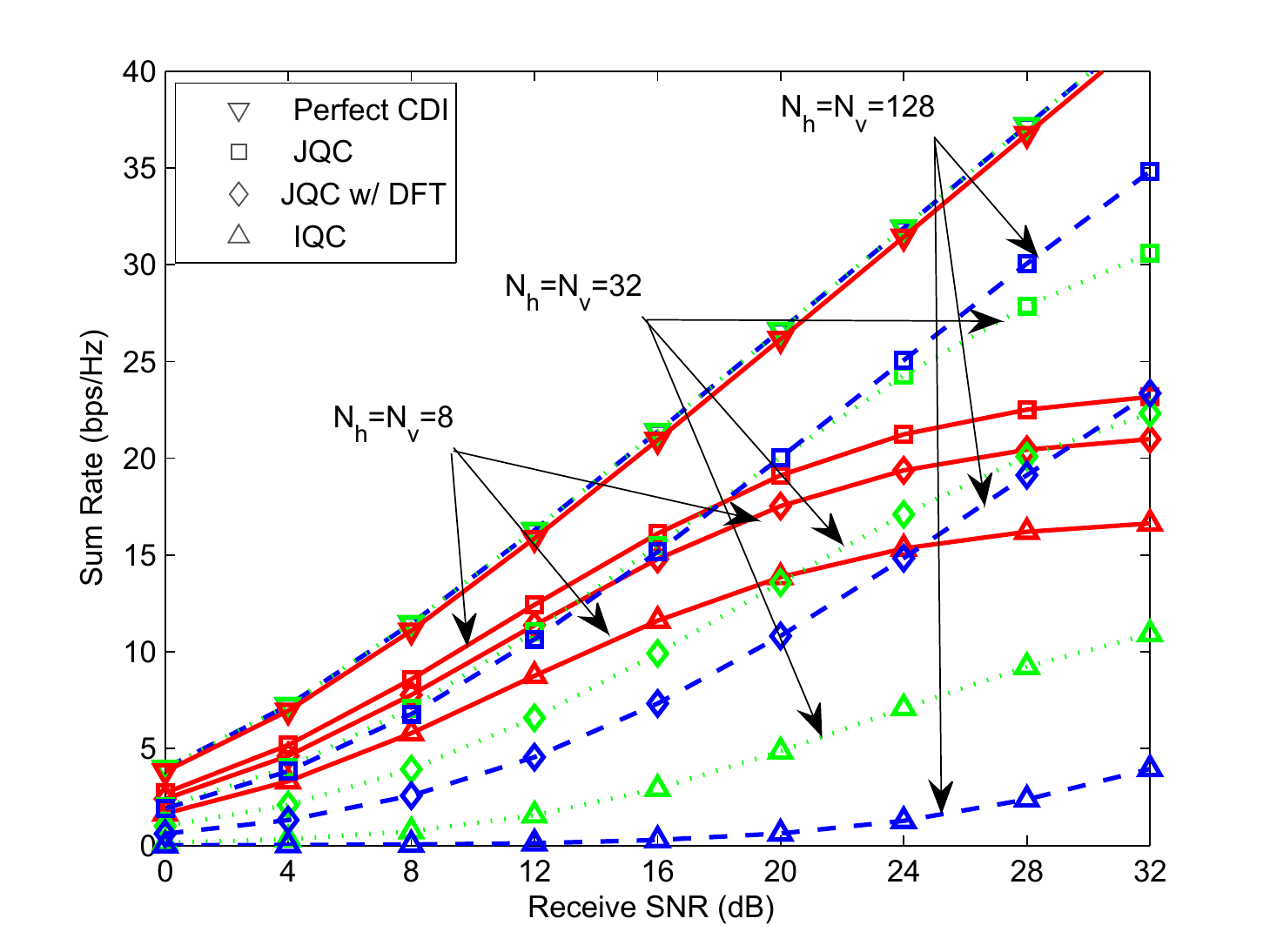}
   \caption{Average sum rates of 3D MIMO systems using URA with different sizes,
   where $K = 4$, $\sigma=20^\circ$, and $B=8$.}\label{DFT}
\end{figure}

The average sum rates of 3D MIMO systems using the URAs with different sizes
are shown in Fig. \ref{DFT}. To show the impact of imperfect statistical information, we also
evaluate the performance of the joint CDI quantization codebook
with quantized statistical information (labeled as ``JQ w/DFT''). The statistical sub-direction information in
horizontal and vertical directions are respectively quantized by a
8-bit DFT codebook \cite{Yang2010}, where each codeword consists of $r_h$ and $r_v$ adjacent  columns in the DFT matrix. The
values of $r_h$ and $r_v$ are selected to quantize the
first $r_h$ and $r_v$ dominant directions such that
$\frac{\sum_{i=1}^{r_h}{{\lambda}_{h,i}}}
{\sum_{i=1}^{N_h}{{\lambda}_{h,i}}}>0.9$ and
$\frac{\sum_{i=1}^{r_v}{{\lambda}_{v,i}}}
{\sum_{i=1}^{N_v}{{\lambda}_{v,i}}}>0.9$, where ${\lambda}_{h,i}$
and ${\lambda}_{v,i}$ are the $i$th element in $\pmb{\lambda}_{h}$
and $\pmb{\lambda}_{v}$ respectively. The power coupling matrix are
quantized by a 8-bit RVQ codebook.

As shown in the figure, when the array size increases, the
performance gaps of the joint CDI quantization codebooks with
perfect and quantized statistical information keep almost constant
under different SNRs and numbers of antennas. This is because the
DFT codewords is a natural choice for representing the statistical
directions for 3D MIMO with URA, as shown in \cite{Adhikary2013}.
Moreover, when the array size increases, the performance gap between
the joint and independent CDI quantization codebook increases. This
is because the rank of channel matrix increases with the size of
antenna array even for a fixed angular spread. It is seen in the
figure, as the size of antenna array increases, the sum rate gap
between perfect CDI and the globally rotated codebook are reduced.
This is because the inter-user interferences are reduced in the
large antenna array when the full spatial correlation information is
available. It indicates the importance of exploiting the channel
correlation information in the codebook design in large array
systems.

\subsubsection{Performance for 3D MIMO Systems with UCCA}
\begin{figure}[t]
\centering
   \includegraphics[width=3.8in]{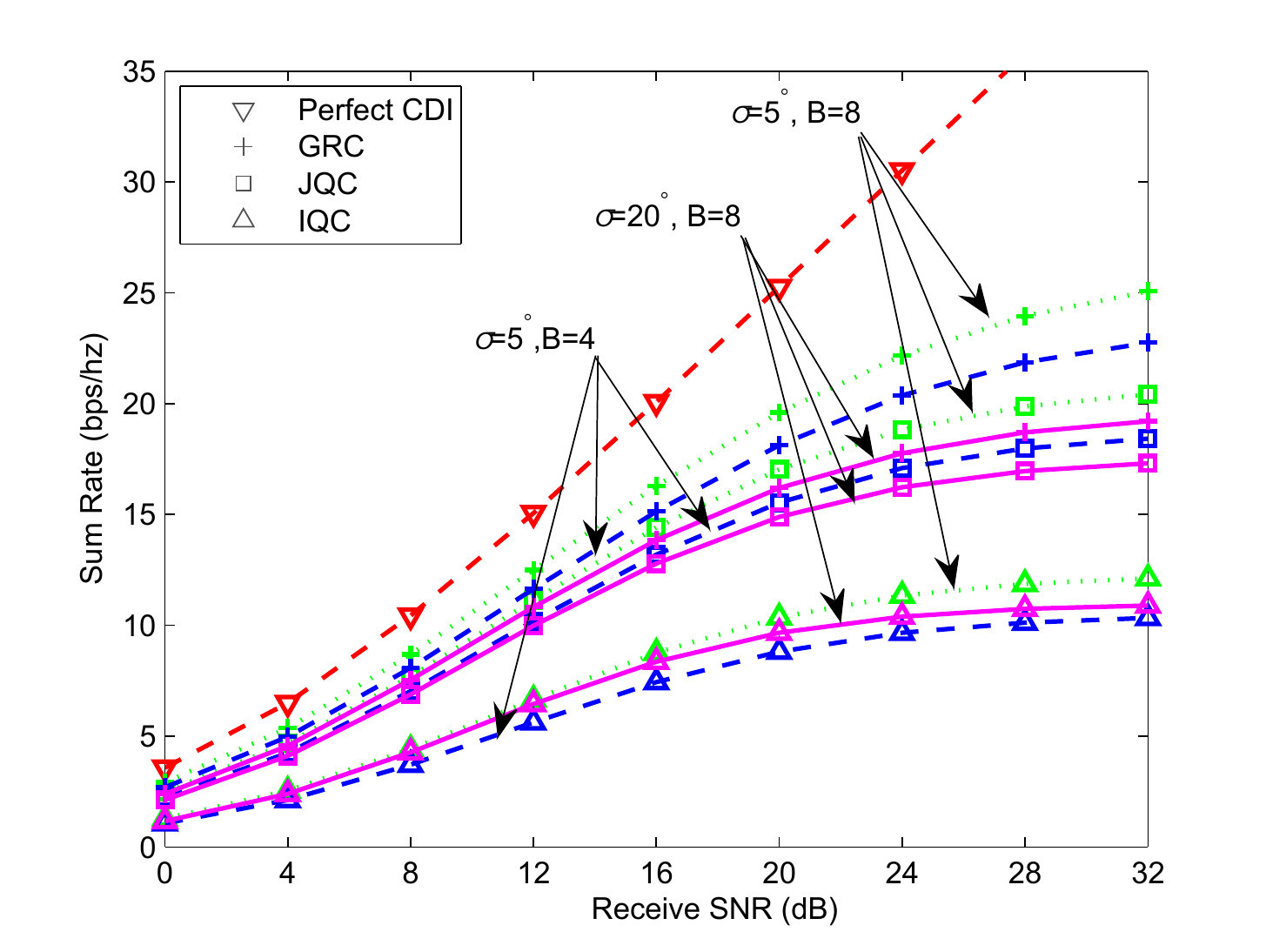}
   \caption{Average sum rates of 3D MIMO systems using UCCAs with different angle spread and number of bits,
   where $K = 4$ and $\sigma=5^\circ$.}\label{UCA}
\end{figure}

We then evaluate the performance of joint CDI quantization codebook
for 3D MIMO systems using the UCCA at the BS. The configuration of
UCCA is given by $J=8$, $L=8$, $d_1=0.5\lambda$, and
$d_{j}=d_{j-1}+0.5\lambda$ for $1<j\leq J$. The average sum rates of
3D MIMO systems are shown in Fig. \ref{UCA}. When obtaining the
statistical information by using the proposed explicit solution, we
set the dimensionality as $N_h=N_v=8$, and similar results can be
observed by other settings.

As shown in the figure, under different angular spreads and numbers of
bits, the performance of joint CDI quantization codebook is
inferior to that of rotated codebook but with acceptable performance
loss. Moreover, the  joint CDI quantization outperforms
the independent quantization significantly.

\subsection{Results in More Realistic Scenarios}
In this subsection, we consider more realistic scenarios in the
simulations. The azimuth and elevation angles in the channel are
modeled by log-normal distributions as in \cite{rel12cm}, where two
scenarios are evaluated: 3D urban micro (UMi) with non line-of-sight
(NLOS), which is referred as 3D UMi, and 3D urban macro (UMa) with
NLOS from an outdoor BS to an indoor user, which is referred as 3D
UMa. The antenna spacing is $0.5\lambda$ for both horizontal and
vertical directions. Unless otherwise specified, the main 3D MIMO
channel parameters are listed in Table I, where $d$ is the distance
from the BS to the user, $h$ is the height of each user. We set $d = 100$ m for 3D UMi, and $d = 250$ m, $h = 1.5$ m for
3D UMa. This channel modeling considers more practical issues, such
as the distance-dependent channel statistics, and the coupling
effect between the statistics like delay spread and angular spread,
which is more complicated than the simplified one in the previous
subsection.

\begin{table*}
 \setbox0\hbox{\verb/\documentclass/}%
  \caption{Main Parameters of 3D MIMO Channels}
   \label{tab1}
    \begin{center}
     \begin{tabular}{c|c|c}
     \hline\hline\\
Scenario & 3D UMi & 3D UMa \\\hline
Mean of azimuth clusters         & $\pmb{U}(-60^\circ,~60^\circ)$ &
$\pmb{U}(-60^\circ,~60^\circ)$
\\\hline Mean of elevation clusters      &
$\pmb{U}(-45^\circ,~45^\circ)$   & $\pmb{U}(-45^\circ,~45^\circ)$
\\\hline
Mean of log delay spread (DS) ($\log_{10}([s])$)   &  -6.89 &
-6.62
\\\hline
Variance of log DS ($\log_{10}([s])$) &  0.54 &  0.32
\\\hline
Mean of log azimuth spread (AS) ($\log_{10}([^\circ])$) & 1.41 &
1.25
\\\hline
Variance of log AS ($\log_{10}([^\circ])$)  & 0.17 & 0.42
\\\hline
Mean of log elevation spread (ES) ($\log_{10}([^\circ])$) &
$\max[-0.5, -2.1(d/1000) +0.9]$ & $\max[-0.5,
-2.1(d/1000)-0.01(h-1.5)+0.9]$
\\\hline
Variance of log ES ($\log_{10}([^\circ])$) & 0.6  & 0.49
\\\hline
Number of clusters &  19 & 12
\\\hline
Number of rays per cluster &  20 &   20 \\\hline\hline
\end{tabular}
 \end{center}
  \end{table*}

\begin{figure*}[t]
\subfigure[3D UMi] {
   \includegraphics[width=3.7in]{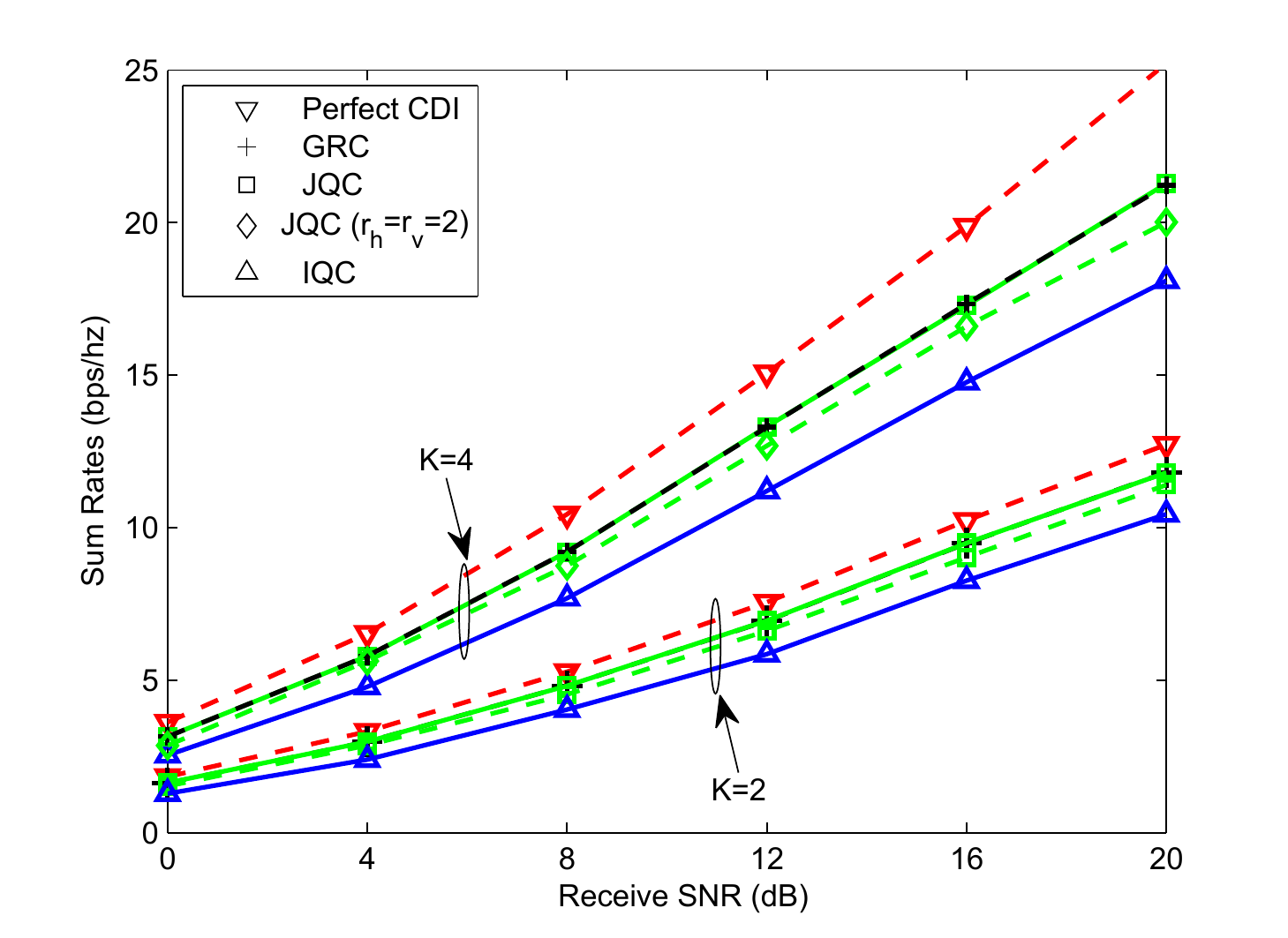}\label{MUrotated}}
\subfigure[3D UMa] {
   \includegraphics[width=3.7in]{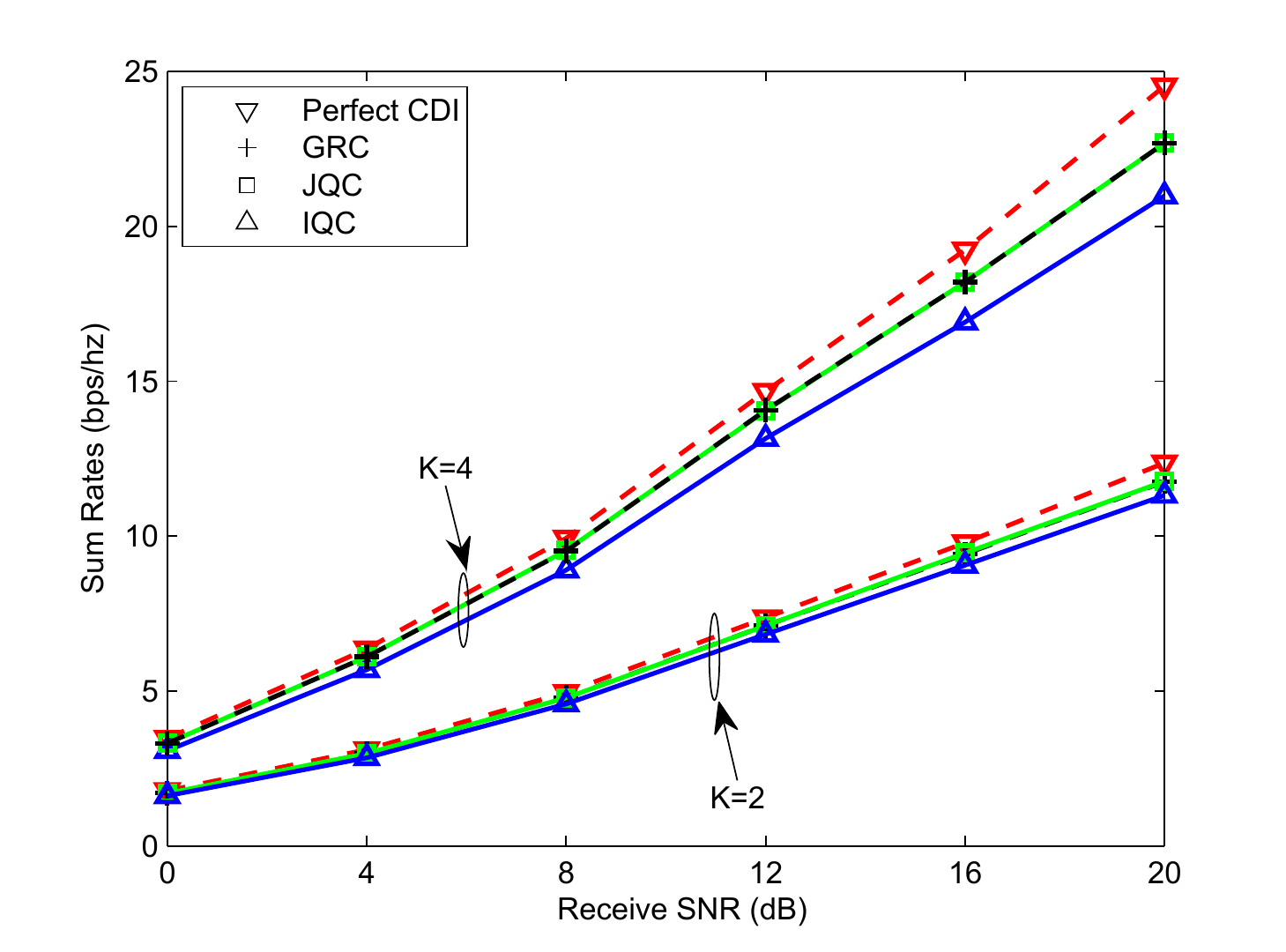}\label{toy4}}
   \caption{Average sum rates of $8\times 8$ 3D  MIMO systems with $B=4$.}\label{sr}
\end{figure*}

We evaluate the average sum rate of an $8 \times 8$ 3D MIMO system
using the joint CDI quantization codebook with different codeword
dimension: JQC with full dimension ($r_h=N_h$ and $r_v=N_v$); JQC
with low dimension ($r_h=2$ and $r_v=2$).



The average sum rates of the 3D MIMO systems with the URA are shown in Fig. \ref{sr}, where  3D UMi and 3D UMa
channels are respectively used. It is shown that the joint CDI quantization codebook with full dimension achieves
the same performance as the globally rotated codebook. Even when using the joint CDI quantization codebook with
reduced quantization dimension, there is still
substantial performance gain over the independent CDI quantization
codebook. In both
scenarios, when the number of users $K$ grows, the
performance gap between the joint and independent CDI quantization
increases. The results are similar to the ones obtained in previous
subsection.

Compared with 3D UMa, the performance gap between the joint and
independent CDI quantization is larger in 3D UMi under the same
number of users. This is because the scattering environment in 3D
UMi is richer than in 3D UMa due to large variance given by the
parameters of log-normal distribution.

\section{Conclusions}
We proposed a joint CDI quantization codebook for 3D MIMO
systems in this paper. By exploiting the unique features of 3D MIMO
channels, the proposed codebook is composed of codewords with a special structure, reflecting different types of information in the 3D MIMO
channel direction. The proposed codebook is easily adapted to different
spatially correlated channels and is applicable to different forms of arrays. Analytical analysis indicated that the proposed
codebook achieves the same performance as the globally rotated codebook for uniform rectangular array under very general channel conditions but
needs much less spatial correlation information. Simulation results
validated the analysis and shown substantial performance gain over independent CDI quantization.

\appendices
\section{Proof of Lemma 1}\label{Proof Lemma1}
For notational simplicity, we express the $i$th column of 3D MIMO
channel matrix $\pmb{H}$ in \eqref{CNM} and the $j$th column of its
transpose $\pmb{H}^T$ as,
\begin{align}
\pmb{c}_i &= \sum\nolimits_{n=1}^N\sum\nolimits_{m=1}^M
g_{n,m}a_{v,i}(\phi_{n,m})
\pmb{a}_{h}(\theta_{n,m}) \nonumber\\
 \pmb{r}_j &=
\sum\nolimits_{n=1}^N\sum\nolimits_{m=1}^M
g_{n,m}a_{h,j}(\theta_{n,m}) \pmb{a}_{v}(\phi_{n,m}) \label{colls}
\end{align}
where $a_{v,i}(\phi_{n,m})$ and $a_{h,j}(\theta_{n,m})$ are
respectively the $i$th entry of vector $\pmb{a}_{v}(\phi_{n,m})$ and
the $j$th entry of $\pmb{a}_{h}(\theta_{n,m})$, $1\leq i \leq N_h$,
$1\leq j \leq N_v$.

We can set $\acute{\pmb{H}}=\pmb{U}_h\pmb{H}_t$, where  $\pmb{U}_h$
and $\pmb{H}_t$ are defined in Proposition 1.

Denote the $i$th column of $\acute{\pmb{H}}$ as
$\acute{\pmb{c}}_{i}$, the $k$th row of $\acute{\pmb{H}}$ as
$\acute{\pmb{r}}_{k}^T$, and the $\ell$th entry of $\acute{\pmb{r}}_{k}$
as $\acute{\pmb{r}}_{k,\ell}$. Then, we have
\begin{align}
\acute{\pmb{r}}_{k}=\pmb{U}_v^H\pmb{r}_{k}\label{rowdefini}
\end{align}

The cross-correlation matrix between the $i$th and $j$th column of
$\acute{\pmb{H}}$ for $i\neq j$ can be obtained as
\begin{align}
\pmb{E}\{\acute{\pmb{c}}_{i}\acute{\pmb{c}}_{j}^H\}
=\pmb{E}\left\{[\acute{\pmb{r}}_{1,i}\cdots\acute{\pmb{r}}_{N_h,i}
]^T[\acute{\pmb{r}}^\ast_{1,j}\cdots\acute{\pmb{r}}^\ast_{N_h,j}]\right\}\label{THs}
\end{align}

For a 3D MIMO system with URA, for any $1\leq k, \ell \leq N_v$ and
$k\neq\ell$, by using \eqref{colls} we have
\begin{align}
E\{\pmb{r}_{k}\pmb{r}_{\ell}^H\}\!&=\!E\Big\{\big(\sum_{n=1}^N\sum_{m=1}^M
g_{n,m}a_{h,k}(\theta_{n,m})
\pmb{a}_{v}(\phi_{n,m})\big)\nonumber\\
&\quad\quad\big(\sum_{n=1}^N\sum_{m=1}^M
g_{n,m}a^\ast_{h,\ell}(\theta_{n,m})
\pmb{a}^{H}_{v}(\phi_{n,m})\big)\Big\}\nonumber\\
&\!=\! \sum_{n=1}^N\!\sum_{m=1}^M\!\!
E\{|g_{n,m}|^2\}E\{a_{h,k}(\theta_{n,m})a^{\ast}_{h,\ell}(\theta_{n,m})\nonumber\\
&\quad\quad \pmb{a}_{v}(\phi_{n,m})
\pmb{a}^{H}_{v}(\phi_{n,m})\}\label{indgandaoa2}\\
&\!=\! \sum_{n=1}^N\!\sum_{m=1}^M\!\!
E\{|g_{n,m}|^2\}E\{a_{h,k}(\theta_{n,m})a^{\ast}_{h,\ell}(\theta_{n,m})\} \cdot \nonumber\\
&\quad\quad\pmb{E}\{ \pmb{a}_{v}(\phi_{n,m})
\pmb{a}^{H}_{v}(\phi_{n,m})\}\label{indgandaoa3}
\end{align}
which are derived under the assumption of weak i.i.d.
rays: \eqref{indgandaoa2} is because the gains $g_{n,m}$ are
uncorrelated with each other and also independent with the angles
$\theta_{n,m}$ and $\phi_{n,m}$; and \eqref{indgandaoa3} is because the
angle $\theta_{n,m}$ is independent with $\phi_{n,m}$.

For i.i.d. variables $\phi_{n,m}$, we have
\begin{align}
&\pmb{E}\{ \pmb{a}_{v}(\phi_{n,m})
\pmb{a}^{H}_{v}(\phi_{n,m})\}
\triangleq   \pmb{R}_v^0=\pmb{U}_v\pmb{\Lambda}_v^{0}\pmb{U}_v^H,
~\text{for}~\forall n ~\text{and}~m\label{iidcora}
\end{align}
where the last equality is the SVD decomposition of positive semi-definite Hermitian correlation matrix
$\pmb{R}_v^0$, and
$\pmb{\Lambda}_v^{0}$ is a positive diagonal matrix.

Then, from \eqref{indgandaoa3} and \eqref{iidcora}, we have
\begin{align}
E\{\pmb{r}_k\pmb{r}_{\ell}^H\}&=\pmb{U}_v\pmb{\Lambda}_v^{k\ell}\pmb{U}_v^H\label{diagnna}
\end{align}
where\begin{align}
\pmb{\Lambda}_v^{k\ell}=\pmb{\Lambda}_v^{0}\sum\limits_{n=1}^N\!\sum\limits_{m=1}^M
\!E\{|g_{n,m}|^2\}E\{a_{h,k}(\theta_{n,m})a^{\ast}_{h,\ell}(\theta_{n,m})\}\nonumber
\end{align}
is diagonal.

From \eqref{rowdefini} and \eqref{diagnna}, we have
\begin{align}
\pmb{E}\{\acute{\pmb{r}}_{k}\acute{\pmb{r}}_{\ell}^H\}=
\pmb{U}_v^H\pmb{E}\{
\pmb{r}_{k}\pmb{r}^H_{\ell}\}\pmb{U}_v=\pmb{\Lambda}_v^{k\ell}
\end{align}
and
\begin{align}
\pmb{E}\{\acute{r}_{k,i}\acute{r}_{\ell,j}^H\}=0,~ \forall k\neq\ell
\text{~and~} i\neq j,\label{THs2}
\end{align}
because the off-diagonal elements of
matrix $\pmb{\Lambda}_v^{k\ell}$ are zero.

From \eqref{THs} and \eqref{THs2}, we obtain
\begin{align}
\pmb{E}\{\acute{\pmb{c}}_{i}\acute{\pmb{c}}_{j}^H\}=\pmb{0}, \forall
i\neq j\label{uncorlt}
\end{align}
which means that the columns of $\acute{\pmb{H}}$ are uncorrelated
with each other. The lemma is proved.

\section{Proof of Lemma 3}\label{Proof Lemma3}
Denote the $n$th column of $\pmb{H}_t$ as
$\pmb{c}_{tn}$. By letting $\pmb{H}_t=\pmb{U}_h^H\acute{\pmb{H}}$,
i.e., $\pmb{c}_{tn}=\pmb{U}_h^H\acute{\pmb{c}}_{n}$, for $n\neq m$
we find that
\begin{align}
\pmb{E}\{\pmb{c}_{tn}\pmb{c}_{tm}^H\}=
\pmb{U}_h^H\pmb{E}\{\acute{\pmb{c}}_{n}\acute{\pmb{c}}^H_{m}\}\pmb{U}_h=\pmb{0},
\forall n\neq m \label{lt1}
\end{align}
where the last equality is from \eqref{uncorlt} in Lemma 1.

Similarly, denote the $n$th row of $\pmb{H}_t$ as $\pmb{r}_{tn}^T$.
By letting $\pmb{H}_t=\grave{\pmb{H}}\pmb{U}_v^\ast$, i.e.,
$\pmb{r}_{tn}=\pmb{U}_v^H\grave{\pmb{r}}_{n}$, for $n\neq m$ we find
that
\begin{align}
\pmb{E}\{\pmb{r}_{tn}\pmb{r}_{tm}^H\}=
\pmb{U}_v^H\pmb{E}\{\grave{\pmb{r}}_{n}\grave{\pmb{r}}^H_{m}\}\pmb{U}_v=\pmb{0},
\forall n\neq m \label{lt2}
\end{align}
where the last equality is obtained from Lemma 2.

From \eqref{lt1} and \eqref{lt2}, the entries in $\pmb{H}_t$ are
uncorrelated. The lemma is proved.

\section{Proof of Theorem}\label{Proof Theorem}
From Lemma 3, the entries of $\pmb{H}_t$ are
uncorrelated. When the spatial correlation matrix of the 3D MIMO channels is perfect, from \eqref{rcr} we have
\begin{align}
\pmb{\hat R}=\pmb{R}&=\pmb{E}\{\text{vec}(\pmb{H})\text{vec}(\pmb{H})^H\}\nonumber\\
&=(\pmb{U}_v\otimes\pmb{U}_h)\text{diag}(\text{vec}\{\pmb{\Lambda}\odot{\pmb{\Lambda}}\})
(\pmb{U}_v\otimes\pmb{U}_h)^H \nonumber
\end{align}

Then, the rotation matrix used in the globally rotated codeword
in \eqref{rote} is
\begin{align}
\pmb{R}^{1/2}=(\pmb{U}_v\otimes\pmb{U}_h)\text{diag}(\text{vec}\{\pmb{\Lambda}\})
\end{align}
which is the same as the rotation matrix used in the joint CDI
quantization \eqref{CJS2} when the statistical information
$\hat{\pmb{U}}_h=\pmb{U}_h$, $\hat{\pmb{U}}_v={\pmb{U}}_v$, and
$\hat{\lambda}=\text{vec}({\pmb{\Lambda}})$ are perfect.

Therefore, when all the instantaneous codewords for the joint CDI
quantization codebook and those for the globally rotated codebook
are identical, the two codebooks achieve the same performance. The
theorem is proved.

\bibliographystyle{IEEEbib}
\bibliography{IEEEabrv,YF_bib}
\end{document}